\documentclass{ws-procs975x65}

\begin{document}

\title{Dynamics in one-dimensional spin systems -- DMRG study }

\author{S. NISHIMOTO}

\address{Max-Planck-Institut f{\"u}r Physik Komplexer Systeme,\\
Dresden, D-01187, Germany\\
E-mail: nishim@mpipks-dresden.mpg.de\\
www.mpipks-dresden.mpg.de}

\author{M. ARIKAWA\footnote{Present address: Institute of Physics, Univ. of Tsukuba 1-1-1 Tennodai, Tukuba Ibaraki 305-8571, Japan}}

\address{Yukawa Institute for Theoretical Physics, Kyoto University,\\
Kyoto, 606-8502, Japan\\
E-mail: arikawa@yukawa.kyoto-u.ac.jp\\
www.yukawa.kyoto-u.ac.jp}

\begin{abstract}
We study the one-dimensional $S=1/2$ Heisenberg model with a uniform 
and a staggered magnetic fields, using the dynamical density-matrix 
renormalization group (DDMRG) technique. The DDMRG enables us to investigate 
the dynamical properties of chain with lengths up to a few hundreds, 
and the results are numerically exact in the same sense as 
'exact diagonalization' results are. Thus, we can analyze the low-energy 
spectrum almost in the thermodynamic limit. In this work, we calculate 
the dynamical spin structure factor and demonstrate the performance of 
the DDMRG method applying the open-end boundary condition as well as  
the periodic boundary condition.
\end{abstract}

\keywords{Dynamical spin structure factor, Heisenberg model, DDMRG, Bethe Ansatz, Magnetic field}

\bodymatter

\section{Introduction\label{sec1}}

In one-dimensional electron systems,
the conformal field theory has succeeded in
the description of the low-energy physics of the Tomonaga-Luttinger liquid.
Recent experimental progress, 
such as angle resolved photoemission and neutron scattering, demands the  
understanding of the elementary excitation pictures of the dynamics
not only in the low-energy and long-wave length limit,
but over the wide range of frequency and momentum. 

In no magnetization case,
M\"{u}ller {\itshape et al}~\cite{mueller} proposed the approximate conjecture of the spin structure factor $S(q,\omega)$ 
in the one-dimensional (1D) $S=1/2$ 
 antiferromagnetic Heisenberg model with nearest neighbor interaction.
By usage of the mathematical works\cite{jimbo}, the exact expression for two-spinon contributions 
 to $S(q,\omega)$ in the thermodynamic limit was found~\cite{karbach}. 
The exact result and M\"{u}ller ansatz for $S(q,\omega)$ have the same singularity at the
lower spectral boundary, which is called the des Cloizeaux-Pearson mode. 
Remarkable progress has been made on the Bethe Ansatz (BA)~\cite{kitanine}, which enables us to numerically calculate the  
$S(q,\omega)$ for the Heisenberg model in a magnetic field for relatively large system size~\cite{biegel}. 

Recently, a dynamical density-matrix renormalization group (DDMRG) method has been developed 
for calculating dynamical correlation functions at zero temperature in quantum lattice models. 
It is based on a variational formulation of the correction vector technique~\cite{kuehner,eric}. 
This method is an extension of the standard DMRG method~\cite{white} which is a powerful 
numerical technique for a variety of 1D systems, whereby we can obtain very accurate ground 
state and low-lying excited states. So far, it has been shown that the momentum-dependent 
dynamical quantities can be calculated with good resolution in finite open Hubbard chains by 
the DDMRG method~\cite{holger}. We could also expect comparable performance for the same model 
in magnetic fields. In this work, we apply the DDMRG method to the 1D $S=1/2$ 
Heisenberg model with an uniform and a staggered magnetic fields and calculate the spin 
structure factor $S(q,\omega)$. The accuracy of the DDMRG method is checked by comparing to 
some analytic solutions.
In addition, the difference of performance by the boundary 
conditions is demonstrated. We note that the Heisenberg model can be dealt with the DMRG 
technique much easier than the Hubbard model.

This paper is organized as follows.
In the next section, we introduce the model and the spin structure factor.
Section \ref{sec3} is devoted to show the brief outline of the DDMRG method.
In Section \ref{sec4}, we demonstrate the DDMRG calculations of the spin structure factor 
for various magnetic fields. The performance with open boundary condition (OBC) as well as 
with periodic boundary condition (PBC) is examined in comparison with the exact solutions. 
Some remarks are given in the final section.

\section{Model\label{sec2}}

We consider the 1D $S=1/2$ Heisenberg model in magnetic fields: 
\begin{equation}
\hat{H} = J \sum_i \vec{S}_i \cdot \vec{S}_{i+1} - H \sum_i S^z_i + h \sum_i (-1)^i S^z_i,
\label{hamiltonian}
\end{equation}
where $J(>0)$ is an antiferromagnetic exchange interaction between neighboring sites. 
Henceforth we take $J$ as the unit of energy. $H$ and $h$ are a uniform and a staggered 
magnetic fields along the $z$-direction, respectively.

The dynamical spin structure factor is defined as 
\begin{equation}
S^{\alpha \beta} (q,\omega) = 
\sum_{n>0}        
\langle \psi_0 | \hat{S}^\alpha_{-q} | \psi_n \rangle 
\langle \psi_n | \hat{S}^\beta_q | \psi_0 \rangle 
\delta(\omega-E_n+E_0),
\label{strucfact}
\end{equation}
where $\hat{S}^\beta_q$ ($=\hat{S}{^\alpha_{-q}}^\dagger$) 
is the Fourier transform of the 
spin operator $\hat{S}^\beta_i$ at site $i$, and $E_n$ and $\left| \psi_n \right\rangle$ are, 
respectively, the $n$-th eigenenergy 
and eigenstate of the system (with the ground state denoted by $n = 0$). Note that the 
definition of the momentum-dependent spin operator $\hat{S}^\beta_q$ depends on choosing 
the boundary conditions. The definitions are given in Section \ref{sec4}.

\section{DDMRG method\label{sec3}}

In this section, the DDMRG method is briefly summarized for calculating the dynamical spin 
structure factor (\ref{strucfact}). 
Thus, we are interested in the momentum-dependent 
spectral functions, 
\begin{eqnarray}
 A(q,\omega)=\lim_{\eta\to 0}\frac{1}{\pi}{\rm Im}
               \langle \psi_0 | 
                 \hat{S}^\alpha_{-q} \frac{1}{\hat{H}+\omega-E_0-i\eta}
                 \hat{S}^\beta_q
               |\psi_0\rangle.
\label{spec}
\end{eqnarray}
The dynamical spin structure factor is then obtained as
\begin{eqnarray}
\nonumber 
 S(q,\omega)&=&-\frac{1}{\pi} {\rm Im} A(q,\omega)\\
 &=& \lim_{\eta\to 0}{\rm Im}
               \langle \psi_0 | 
                 \hat{S}^\alpha_{-q} \frac{\eta}{(\hat{H}+\omega-E_0)^2+\eta^2}
                 \hat{S}^\beta_q
               |\psi_0 \rangle.
\label{spec2}
\end{eqnarray}
To compute the dynamic quantity such as Eq.~(\ref{spec}), we use the DDMRG method. This 
approach is based on a variational principle. One can easily show that for $\eta > 0$ and fixed frequency $\omega$ the minimum of the functional
\begin{equation}
W(\psi) =
\langle \psi | (E_0+\omega-\hat{H})^2 +\eta^2 | \psi \rangle \\
+ \eta \langle \psi_0 | \hat{S}^\alpha_{-q} | \psi \rangle
+ \eta \langle \psi | \hat{S}^\beta_q | \psi_0 \rangle
\label{functional}
\end{equation}
with respect to all quantum states $|\psi\rangle$ is
\begin{equation}
W(\psi_{\rm min}) = 
\langle \psi_0 | \hat{S}^\alpha_{-q} 
\frac{-\eta^2}{(E_0+\omega-\hat{H})^2 +\eta^2}
\hat{S}^\beta_q | \psi_0 \rangle.
\end{equation} 
The functional minimum is related to the convolution of the dynamical spin structure 
factor~(\ref{spec2}) with a Lorentz distribution of width $\eta$ by
\begin{equation}
W(\psi_{\rm min}) = -\pi \eta S(q,\omega).  
\end{equation}

The DDMRG method consists essentially of  
minimizing the functional~(\ref{functional}) numerically using
the standard DMRG algorithm. 
Thus the DDMRG provides the dynamical spin structure factor 
$S^\eta(q,\omega)$ for a finite broadening $\eta$.
The full spectrum~(\ref{spec2}) convolved with the Lorentz distribution is given as
\begin{equation}
S^{\eta}(q,\omega) =  \int_{-\infty}^{\infty} {\rm d}\omega^\prime
S(q,\omega^\prime) \frac{\eta}{\pi[(\omega-\omega^\prime)^2+\eta^2]}
\label{convolution}
\end{equation}
The real part of Eq.~(\ref{spec}) can be calculated 
with no additional computational cost but is generally less accurate.
The necessary broadening
of spectral functions in DDMRG calculations
is actually very useful for studying continuous spectra
or for doing a finite-size scaling analysis~\cite{eric}.

If one would like to obtain the spectrum in the limits of $L \to \infty$ and $\eta=0$, 
it can be done by carrying out a deconvolution of the DDMRG data~\cite{nishim}. 
In theory, the deconvolution scheme corresponds to solving (\ref{convolution}) for obtaining 
$S_(q,\omega)$, where a set of $S^\eta_(q,\omega)$ on the left-hand side is calculated 
with the DDMRG method.~\cite{nishim}. We also know that the broadened spectrum of the 
system on an infinite lattice ($L \to \infty$) is usually almost identical to the spectrum 
on a finite lattice ($N < \infty$) if $\eta \ge c/L$ (the constant $c$ is comparable to 
the width of the spectrum). Therefore, assuming that the DDMRG data $S^\eta(q,\omega)$ 
describes the broadened spectrum for $N \to \infty$, one can solve (\ref{convolution}) 
approximately under the condition that $S(q,\omega)$ is the convolved `exact' spin structure 
factor. We note that $S^\eta(q,\omega)$ must be a continuous and relatively smooth function 
in order to obtain quantitatively accurate spectrum after deconvolution. To this end, it is 
required to choose $\eta$ smaller than the width of the spectrum in $\eta=0$.~\cite{eric} 

\section{Results\label{sec4}}

\subsection{Field-induced magnetization}

\begin{figure}[t]
\begin{center}
\psfig{file=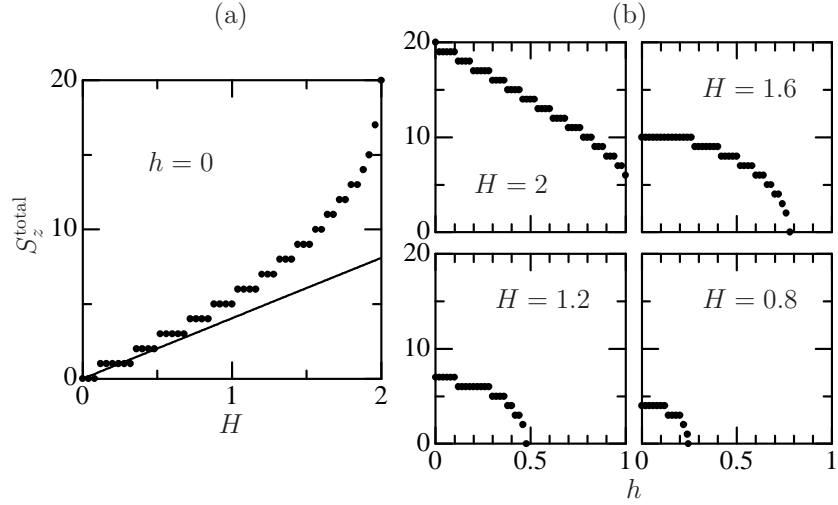,width=11.0cm}
\end{center}
\caption{$z$-component of the total-spin quantum number as a function of $H$ (a), and as a 
function of $h$ for various fixed $H$ (b). The solid line denotes the exact behavior for 
$H \to 0$.}
\label{totalSz_uni}
\end{figure}

The $z$-component of the total-spin $S_z^{\rm tot}$ is now a good quantum number for 
the DMRG calculation. Therefore, it is very efficient to classify the Hamiltonian by $S_z^{\rm tot}$. 
Let us first study $S_z^{\rm tot}$ of the ground state in the system with magnetic fields 
before starting the dynamical calculations. In Fig.~\ref{totalSz_uni}(a), we show $S_z^{\rm tot}$ 
of the ground state as a function of the uniform magnetic field $H$. The calculations are 
carried out with the DMRG method in a chain with $L=40$ sites under the PBC. Note that 
$S_z^{\rm tot}$ is equivalent to the magnetization of the system via the relation 
$M=S_z^{\rm tot}/L$. 

For small $H$, the magnetization is induced proportionately with the uniform field, 
$M=(2/\pi^2)H$.~\cite{griffiths} The derivative $\frac{\partial M}{\partial H}$ becomes larger 
with increasing $H$ since the antiferromagnetic spin fluctuations are rapidly weakened with 
approaching to the fully-polarized phase. The system is fully polarized at a critical uniform field 
$H_{\rm c}=2$ where the spin susceptibility diverges, i.e., 
$\frac{\partial M}{\partial H} \to \infty$. We also show the dependence of $S_z^{\rm tot}$ on 
the staggered magnetic field $h$ for several kinds of $H$ values in Fig.~\ref{totalSz_uni}(b). 
We can see that the magnetization is rapidly suppressed by the staggered field and 
it reaches to zero at a critical value $h_{\rm c}$. The critical staggered field is roughly scaled 
as $h_{\rm c} \sim H/2$.

\subsection{Dynamical spin structure factor}

\subsubsection{Periodic boundary condition}

Normally, the OBC is applied to the (D)DMRG calculations in order to achieve an accurate 
treatment; namely, we can study systems with the OBC on much larger lattices than systems 
with the PBC, with keeping a sufficient accuracy. Especially in interacting fermion systems such 
as the Hubbard and the $t$$-$$J$ models, the OBC seems to be rather essential to adequately 
truncate their large Hilbert space. However, a balk system is generally better described using 
the PBC, and moreover the exact definition of the momentum-dependent operators with the OBC 
is quite difficult.

\begin{figure}[t]
\begin{center}
\psfig{file=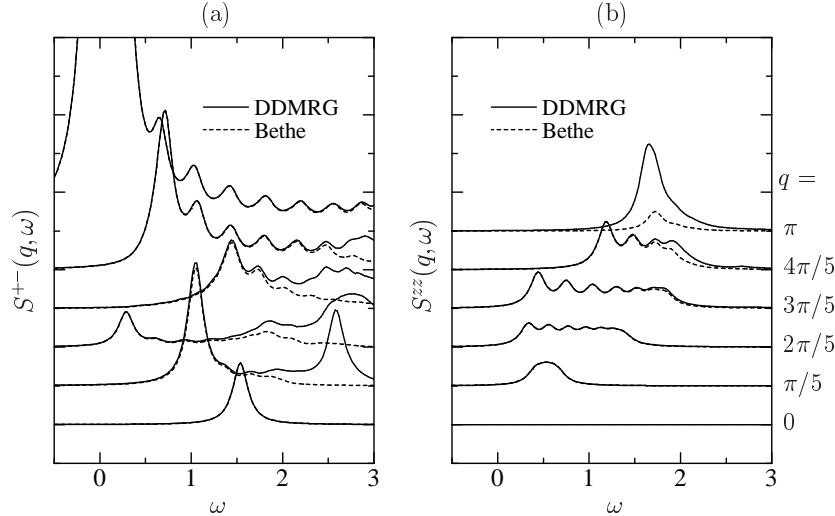,width=11.0cm}
\end{center}
\caption{Spin structure factor $S^{+-}(q,\omega)$ (a) and $S^{zz}(q,\omega)$ (b) for $L=40$, 
$H=1.537$, and $\eta=0.1$ applying the PBC. Dotted curves show the BA solution with 
$\eta=0.1$ for comparison: up to two-spinon and two-antispinon contributions (a) and 
up to two-spinon and one-antispinon contributions (b). 
}
\label{cf_Bethe}
\end{figure}

Here, an idea will naturally arise as follows: it may be that 
the PBC can be applicable in a spin system where the Hilbert space is much more reduced than 
that in the fermion systems. Motivated by this idea, we check the DDMRG performance with 
the PBC in the 1D $S=1/2$ Heisenberg model. 
We now start to calculate the dynamical spin structure factor (\ref{strucfact}). 
With the PBC, the spin operators $S^\alpha_q$ are defined by
\begin{equation}
S^\alpha_q = \frac{1}{\sqrt{L}} \sum_l e^{iql} S^\alpha_l,
\label{operator_pbc}
\end{equation}
with momentum $q=2\pi z/L$ for integers $-L/2 < z \le L/2$.

Using BA solutions on a finite lattice, the spectral weight are
available in the case with the uniform magnetic field~\cite{kitanine,biegel}. Comparison 
with the BA solution will provide an opportunity to test the performance of the DDMRG method. 
In Fig.~\ref{cf_Bethe}, we compare the DDMRG spectra with the line shapes based on the BA 
solutions in $H=1.537$, where the system size is $L=40$ sites and the broadening is taken 
as $\eta=0.1$ for both of the methods. We can see 
excellent agreements in the low-energy excitations, whereas the BA spectra deviate 
from the DDMRG data in the high-frequency range. It means that the DDMRG method can take 
into account higher-order terms than the BA treatment. If the discarded weight 
$w_{\rm d}$ is sufficiently small in the DMRG truncation procedure, e.g., $w_{\rm d}<10^{-5}$, 
the DDMRG results for finite systems are numerically exact in the same sense as `exact 
diagonalization' results are. In the present calculations, we keep up to $m=300$ density-matrix 
eigenstates and the typical discarded weight is less than $5 \times 10^{-6}$. Thus, we are 
confident that the DDMRG method using the PBC works well in the $S=1/2$ Heisenberg 
chain with at least up to several tens of sites.

\begin{figure}[t]
\begin{center}
\psfig{file=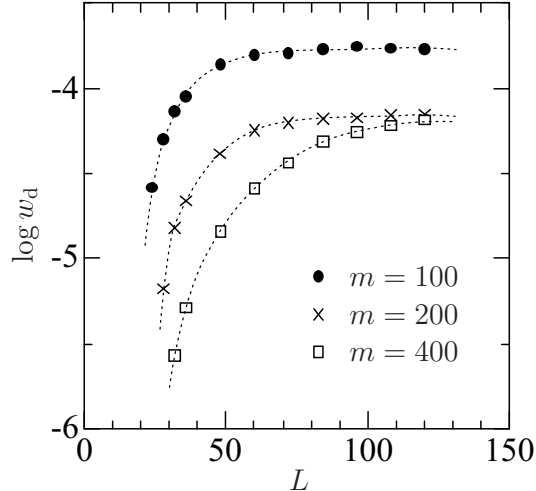,width=7.0cm}
\end{center}
\caption{Average discarded weight $w_{\rm d}$ as a function of system size $L$ for the truncation number
$m=100$, $200$, and $400$ in the calculation of $S^{+-}(0,\omega)$. Dotted lines are guides 
for eyes.
}
\label{disc_weight}
\end{figure}

Let us now mention how large system can be dealt with using the DDMRG method. The 
performance depends mainly on the number of the density-matrix eigenstates truncated $m$. 
As the truncation number $m$ increases, the accuracy of calculation is improved but more CPU time and disk space 
are required. Practically, we could keep a several hundreds density-matrix eigenstates with 
the present typical cluster machine, e.g., Pentium 4 3.2GHz or Opteron 252 2.6GHz. 
In Fig.~\ref{disc_weight}, we show the logarithm of average discarded weight $w_{\rm d}$ 
as a function of system size $L$ for $m=100$, $200$, and $400$ in the calculation of 
$S^{+-}(0,\omega)$. Naturally, $w_{\rm d}$ decreases as $m$ increases at a fixed system size. 
On the other hand, for a fixed $m$, $w_d$ increases rapidly with increasing $L$ for smaller 
systems and is almost constant for larger systems. Empirically, we can no longer expect a good DMRG performance in this `constant' 
region. For example, we estimate $L=40$, $60$, and $80$ as the optimum system sizes for 
$m=100$, $200$, and $400$, respectively. Therefore, the DDMRG calculation with about a hundred 
lattice sites would be adequately possible in the standard $S=1/2$ Heisenberg chain using the PBC 
if we take $m \approx 600$. 

\begin{figure}[t]
\begin{center}
\psfig{file=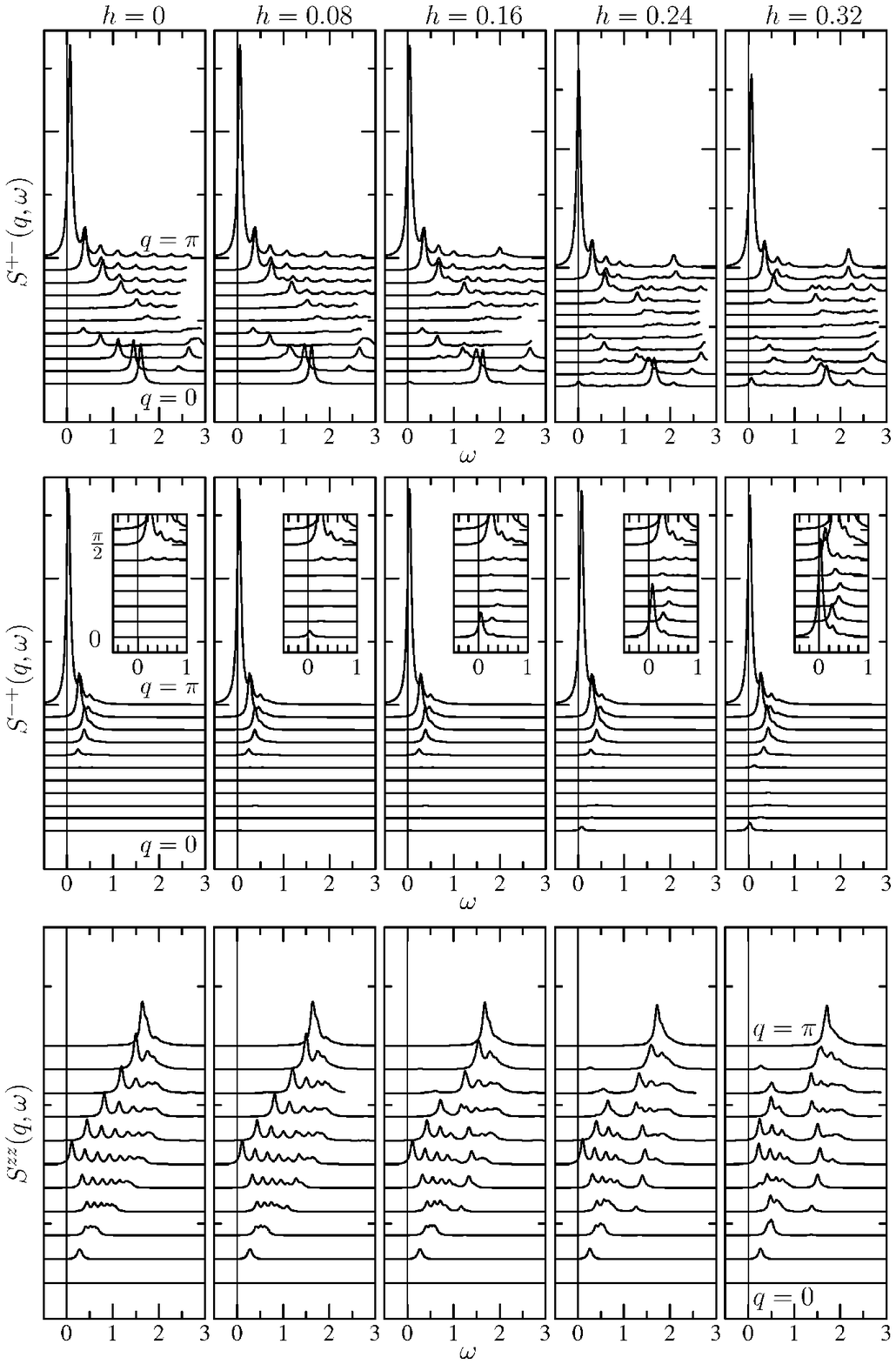,width=10.0cm}
\end{center}
\caption{Spin structure factor $S^{+-}(q,\omega)$ (upper), $S^{-+}(q,\omega)$ (middle), and 
$S^{zz}(q,\omega)$ (bottom) calculated with the DDMRG method for $L=40$ and $\eta=0.05$ 
applying the PBC. The staggered magnetic field $h$ is varied with fixing the uniform 
magnetic field at $H=1.6$.
}
\label{Sqw_L40}
\end{figure}

We then add the staggered magnetic field to the system under a fixed uniform magnetic field.  
In Fig.~\ref{Sqw_L40}, we show the spin structure factors $S^{\alpha \beta}(q,\omega)$ for 
$\alpha \beta = +-, -+, zz$ at $h=0$, $0.08$, $0.16$, $0.24$, and $0.31$ with fixing $H=1.6$, 
calculated with the DDMRG method for $L=40$ and $\eta=0.05$ applying the PBC. 
The high-energy structures, which cannot be obtained by the conformal field theory, 
are clearly seen. 

For $h=0$ [see also Fig.~\ref{cf_Bethe}], the feature of $S(q,\omega)$ is very similar to 
that of Haldane-Shastry model ($1/r^2$ interaction)~\cite{haldane,shastry,zirnbauer,talstra}. 
In the lower energy edge of the small $q$ ($0 \le q < 2 \pi S_z^{\rm tot}/L$) in the 
$S^{+-}(q,\omega)$, the one antispinon (magnon) contributes as a $\delta$-function peak. 
At $q=0$, the exact form is known, $S^{+-}(0,\omega)=(4\pi S_z^{\rm tot}/L)\delta(\omega-H)$, 
which presents a resonant mode of the magnetization induced by the uniform magnetic 
field.\footnote{The one-antispinon and the two-spinon plus two-antispinon contribution to 
$S^{+-}(q,\omega)$ in Haldane-Shastry model can be expressed in terms of the analytic 
expression of the advanced Green function for the spinless Sutherland model with coupling 
parameter $\lambda=2$.~\cite{serban}. Other components, $S^{-+}(q,\omega)$ and the two-spinon 
plus one-antispinon contribution to $S^{zz}(q,\omega)$ for Haldane-Shastry model 
also can be expressed by the correlation function of the spinless Sutherland 
model~\cite{saiga, arikawa06}.
}
The spectra are changed gradually with increasing $h$ and we can see some predominant features 
for $h=0.32$ as the following:
\begin{itemlist}
\item The low-energy structures of $S^{+-(-+)}(q,\omega)$ around $q=0$ arise. They 
are associated with increase of the weight of unfluctuating antiferromagnetic configuration 
in the ground state.
\item The one-magnon peaks of $S^{+-}(q,\omega)$ in the small momentum region 
are smeared and they will disappear when $h > h_{\rm c}$.
\item In the large momentum ($q \sim \pi$) region of $S^{+-}(q,\omega)$, the continuum structure 
when $h=0$ turn to a few discrete peaks due to the decrease of spin fluctuations.
\item The band structure of $S^{zz}(q,\omega)$ seems to be split into two small-dispersive 
bands around $\omega \lesssim 0.5$ and $\omega \sim H$.
\end{itemlist}

\subsubsection{Open-end boundary condition}

\begin{figure}[t]
\begin{center}
\psfig{file=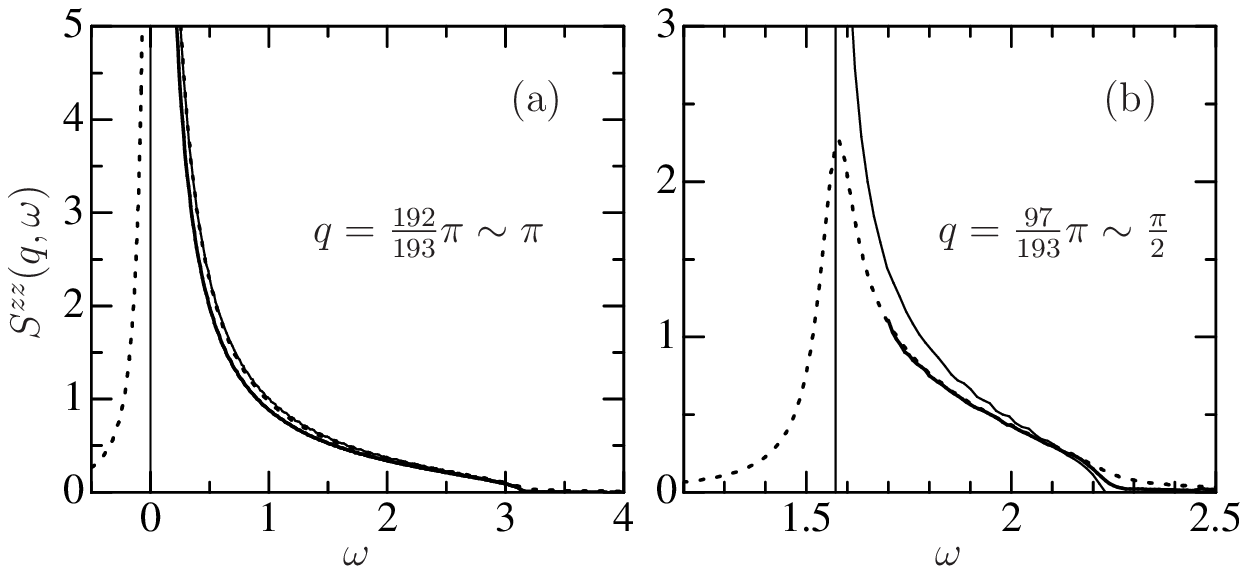,width=10.0cm}
\end{center}
\caption{Spin structure factor $S^{zz}(q,\omega)$ in a chain with $L=192$ sites for 
$q=(192/193)\pi \sim \pi$ (a) and $q=(97/193)\pi \sim \pi/2$ (b), calculated with the DDMRG 
method for $\eta=0.05$ (dotted lines) applying the OBC and then deconvoluted (bold lines). 
The thin lines show the analytic solutions in the thermodynamic limit~\cite{karbach}. 
}
\label{nomag_L192}
\end{figure}

We now turn to the calculation with the OBC. A DDMRG result with the OBC is less exact 
than that with the PBC for the same system size. However, the OBC enable us to carry out 
a calculation for much larger systems, and then we may obtain the dynamical quantities in the 
limits of $L \to \infty$ and $\eta \to 0$ through additional techniques such as the finite-size 
scaling or the deconvolution techniques. Hence, it would be also useful to check a performance 
with the OBC.

When the OBC is applied, we usually use the eigenstates of the particle-in-a-box problem to 
define the operators 
\begin{equation}
S^\alpha_q = \sqrt{\frac{2}{L+1}} \sum_l \sin(ql) S^\alpha_l,
\label{operator_obc}
\end{equation}
with quasi-momentum $q=\pi z/(L+1)$ for integers $1 < z \le L$. This definition of $S^\alpha_q$ 
should be equivalent to that for the PBC in the thermodynamic limit, but the 
%equivalence 
agreement may 
not necessarily good in finite systems. In Fig.~\ref{nomag_L192}, we show the DDMRG results of 
$S^{zz}(q,\omega)$ without magnetic field for $L=192$ and $\eta=0.05$ with OBC, as well as 
the exact line shape of the two-spinon excitation contribution for $L \to \infty$~\cite{karbach}. 
In order to compare the finite-size DDMRG spectrum with the 
%BA 
exact solution without broadening, 
we need to eliminate the broadening $\eta$ from the `convolved' DDMRG spectrum. 

An approach for obtaining the spectrum in the $\eta=0$ limit is the deconvolution of the DDMRG 
spectrum. Theoretically, a deconvolution means solving (\ref{convolution}) for $S(q,\omega)$ 
using the DDMRG data in the left-hand side. If it was possible to do this calculation exactly, 
one would obtain the discrete spectrum on a finite lattice of $L$ sites. Nevertheless, the 
broadened spectrum on a infinite lattice ($L \to \infty$) is usually almost identical to the 
discretized DDMRG spectrum ($L < \infty$) under the condition $\eta \ge c/L$. Since the width 
of $S^{zz}(q,\omega)$ is always less than $c=4$ according to the 
%BA 
exact solution, our choice 
$\eta=0.05$ indeed satisfies the condition. 

At $q=(192/193)\pi$, the deconvoluted DDMRG data seems to agree well with the exact solution 
in the presented scale [see Fig.~\ref{nomag_L192} (a)]. Actually, the deconvoluted DDMRG data 
is a shade thinner than the exact solution. It might come from the difference of momentum taken 
in the calculations, i.e., $q=(192/193)\pi$ in the DDMRG and $\pi$ in the 
%BA calculations
exact solutions. 
Nevertheless, it can be rather hard to find the existence of $1/\omega$-divergence for the 
lower edge $\omega=0$ expected from the exact solution. If the width of a peak for $\eta=0$ is 
smaller than $\eta$, the we cannot obtain it accurately by the deconvolution technique. 
Hence, we could not see the very sharp structure of $S^{zz}(\pi,\omega)$ at $\omega \le 0.05$ 
correctly. On the other hand, the agreement is not very good at $q=(97/193)\pi \sim \pi$ 
[see Fig.~\ref{nomag_L192} (b)]. Occasionally, quantitative estimation would be somewhat problematic. We find that the accuracy of spectrum is usually the worst around $q=\pi/2$ 
for using the OBC. However, we suggest that the definition of operators with the quasi-momentum 
gives a good approximation at least qualitatively.

\begin{figure}[t]
\begin{center}
\psfig{file=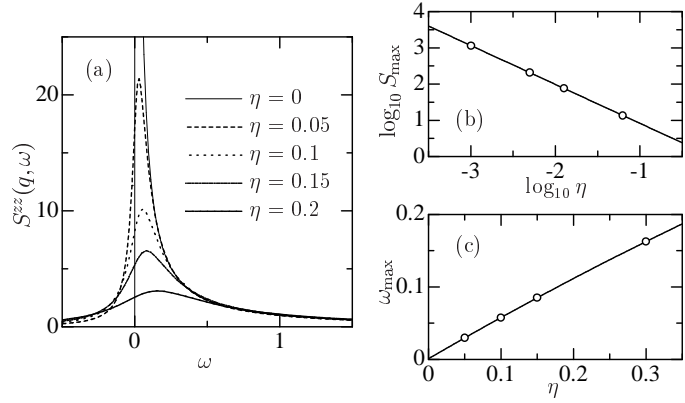,width=9.0cm}
\end{center}
\caption{(a) Spin structure factor $S^{zz}(q,\omega)$ at $q=(L-1/L)\pi \sim \pi$ for various 
system lengths $L$ and $\eta=9.6/L$. (b) Extrapolation of $S_{\rm max}$ to $\eta \to 0$. 
(c) Extrapolation of $E_{\rm max}$ to $\eta \to 0$.}
\label{nomag_scaling}
\end{figure}

The other approach is the finite-size scaling analysis of the DDMRG data.~\cite{eric,holger} 
In Fig.~\ref{nomag_scaling} (a), we show the spin structure factor $S^{zz}(q,\omega)$ at 
$q=(192/193)\pi$ for several system sizes $L$. We keep a relation $\eta L = 9.6$ 
for systematic extrapolation. We can clearly see the convergence of the finite-size spectra 
toward the exact spectrum as $\eta$ decreases. We also can study the lower-edge behavior 
more quantitatively by a scaling analysis of the maximal value in the DDMRG spectrum 
$S^{zz}(q,\omega)$. For instance, the height of the low-energy maximum $\log_{10} S_{\rm max}$ 
is scaled linearly as a function of $\log_{10} \eta$ [see Fig.~\ref{nomag_scaling} (b)] 
and the slope is $-1$. It means that the spectrum diverges as $\eta^{-1}$ for $\eta \to 0$. 
Moreover, the position of the low-energy maximum, $E_{\rm max}$, approaches to $0$ as $\eta$ 
goes to $0$. We can then confirm that $S^{zz}(\pi,\omega)$ has a singularity with exponent $-1$ 
at $\omega=0$ in the thermodynamic limit $L \to \infty$. This is consistent with the exact 
result which shows the $1/\omega$-divergence of $S^{zz}(\pi,\omega)$ at the lower edge 
$\omega=0$.

\section{Remarks and Outlook}

In this work, we demonstrate successful application of both the OBC and the PBC to 
the DDMRG calculation in the 1D $S=1/2$ Heisenberg model. Each of the boundary 
conditions has advantages and disadvantages, and then we need to make proper use of 
the boundary conditions depending on the situation. 
\begin{description}
\item[(i)] With the PBC, the system size must be restricted practically up to about 
a hundred but the result is numerically exact. Therefore, the PBC is suited to study
\begin{description}
  \item[-] quantitative estimation of the whole spectrum, and
  \item[-] accurate picture of complex dispersive structure.
\end{description}
\item[(ii)] With the OBC, we can set the system size to be two hundreds or more, and thus we can 
easily obtain the spectrum in the thermodynamic limit via the finite-size scaling or the 
deconvolution techniques. However, quantitative accuracy is occasionally missing on a finite 
lattice. Therefore, the OBC is suited to study
\begin{description}
  \item[-] the behavior and position of a singularity in the spectrum, and
  \item[-] rough picture of simple dispersive structure.
\end{description}
\end{description}
Possibly, the finite-size scaling or the deconvolution techniques will be applicable to 
the DDMRG data with the PBC if the spectrum had relatively simple form.

Finally, we add that the method can easily be extended to other systems: for example, 
\begin{description}
  \item[-] $S=1/2$ Heisenberg ladder systems,
  \item[-] more realistic models which include the transverse staggered magnetic field 
or the Dzyaloshinsky-Moriya interaction, and
  \item[-] spin systems where the total spin is more than $S=1/2$, etc.
\end{description}

\end{document}